# MODELING OF ION-IMPLANTED ATOMS DIFFUSION DURING THE EPITAXIAL GROWTH OF THE LAYER

O.I. Velichko, V.A. Burko

Department of Physics, Belarusian State University of Informatics and Radioelectronics, 6, P.Brovki Street, Minsk, 220013, Belarus, phone (+37529) 6998078, e-mail: velichkomail@gmail.com

The equation of impurity diffusion due to formation, migration, and dissolution of the pairs "impurity atom — intrinsic point defect" taking into account the nonuniform distributions of nonequilibrium point defects and drift of the pairs in the field of elastic stresses is presented in the coordinate system associated with the moving surface of the growing epitaxial layer. The analytical solution of this equation for the low fluence ion implantation has been obtained.

**Introduction**

One of the basic trends of modern electronics is the wide use of different sandwich structures such as $Si/Si_{1-x}Ge_x$ [1-3]. The semiconductor layer formed can be doped with an impurity during either the epitaxial growth or by method of ion implantation. Then, during the following epitaxy impurity diffusion occurs in the growing layer. It follows from the experimental data that the impurity redistribution is influenced in the latter case by nonequilibrum defects created by ion implantation and by internal elastic stresses arising due to lattice mismatch [1,2]. A phenomenon of impurity segregation on the surface of the growing layer also can play an important role in dopant redistribution [1,2]. For description of impurity atoms transport under consideration it is possible to use the diffusion equation obtained in [4] as this equation takes into account the nonuniformity of distributions of intrinsic point defects (IPD), and drift of the pairs "impurity atom — IPD" in the field of elastic stresses. However, the numerical solution of this equation is complicated by the presence of a moving surface boundary. Therefore, it is reasonable to introduce a new coordinate system $x = x^* - v_B t$ bound to the surface of a growing layer and solve the diffusion equation in this coordinate system. Here $x^*$ is the coordinate counted off the initial surface of the semiconductor (the epitaxial layer is growing on this surface); $v_B$ is the projection of the growth velocity on the $x^*$ axis. Let us assume that the transient enhanced diffusion of ion-implanted impurity occurs for the most part due to the interaction with silicon self-interstitials (I) and we can neglect the formation of vacancy-impurity pairs. Then, the equation for impurity diffusion [4] in the moving coordinate system takes the following form

$$\frac{\partial C}{\partial t} = \frac{\partial}{\partial x}\left\{ D\left[\frac{\partial (\tilde{C}^\times C)}{\partial x} + \frac{\tilde{C}^\times C}{\chi}\frac{\partial \chi}{\partial x}\right]\right\} - \frac{\partial}{\partial x}\left(v_x^F \tilde{C}^\times C\right) + v_B \frac{\partial C}{\partial x} - S^A + G^A . \quad (1)$$

Here $C$ is the concentration of substitutionally dissolved impurity atoms; $\chi$ is the concentration of charge carriers (electrons or holes) normalized to the intrinsic carrier concentration $n_i$; $D = D(\chi)$ is the effective diffusivity of impurity atoms due to formation, migration, and dissolution of the pairs "impurity atom — silicon self-interstitial", $\tilde{C}^\times$ is the concentration of silicon self-interstitials in the neutral charge state normalized to the thermal equilibrium concentration of this species; $v_x^F$ is the projection of the velocity of the drift of the pairs in the field of elastic stresses to the $x$ axis; $S^A$ is the rate of the capture of substitutionally dissolved impurity atoms by extended defects or by clusters of impurity atoms; $G^A$ is the rate of the transfer of impurity atoms into the substitutional position due to annealing of extended defects or cluster dissolution.

It is clear from (1) that the diffusion equation obtained allows to describe the "uphill" diffusion of impurity atoms in the case of nonuniform distribution of silicon self-interstitials in the neutral charge state or due to the drift of the pairs in the field of elastic stresses. However, the numerical solution of the equation (1) is very complicated and, therefore, we need a method to verify the correctness of the numerical results obtained.

The purpose of the paper is to construct the analytical solution of the equation (1) in the simplified case of low fluence ion implantation available for the verification of the numerical solution.

**Boundary value problem of diffusion**

In the case of low fluence ion implantation the impurity concentration $C(x,t) \sim n_i$, and $\chi \approx 1$. It means that impurity diffusivity is independent on the impurity concentration and the drift of charged particles in the built-in electrical field is negligible. Besides, there are no impurity clustering and no capture of impurity atoms by extended postimplantation defects. The concentration of intrinsic defects is near the thermal equilibrium ($\tilde{C}^\times \approx 1$). Let us also to assume that the drift velocity of the pairs $v_x^F$ is constant. Then, the equation (1) takes the following form

$$\frac{\partial C}{\partial t} = D\frac{\partial^2 C}{\partial x^2} + v_{BX}\frac{\partial C}{\partial x} , \quad (2)$$

where $v_{BX} = v_B - v_x^F$.

The boundary value problem under consideration includes the equation of diffusion (2), boundary conditions on the interval $x \in [0, +\infty]$, and the initial condition. To solve the problem a zero Dirichlet boundary conditions on the interface "the growing layer — epitaxial ambient" and in the bulk of the semiconductor have been imposed

$$C(0,t) = 0 \; , \qquad C(+\infty, t) = 0 \; . \qquad (3)$$

As the initial condition we choose the distribution of impurity atoms which is formed by ion implantation and can be presented as Gaussian distribution. Taking into account that $x^* = x$ in the epitaxy beginning, we shall describe the initial distribution by the following expression

$$C(x,0) = C_m \exp\left[-\frac{(x-R_p)^2}{2\Delta R_p^2}\right] \; , \qquad (4)$$

where $C_m$ is the impurity concentration in a maximum of Gaussian distribution; $R_p$ is the average projective range of impurity ions; $\Delta R_p$ is the straggling of the average projective range.

### The analytical solution

To obtain the analytical solution of the formulated boundary value problem, the Smoluchowski substitution is used

$$C(x,t) = \tilde{C}(x,t) \exp\left(-\frac{v_{BX} x}{2D} - \frac{v_{BX}^2 t}{4D}\right) \; , \qquad (5)$$

which reduces the equation (2) to the equation of the Fick's second law with the constant diffusivity written for the function $\tilde{C}(x,t)$

$$\tilde{C}'_t = D \tilde{C}''_{xx} \; . \qquad (6)$$

The boundary conditions (3) and initial condition (4) written for the equation (6) take the following form

$$\tilde{C}(0,t) = 0 \; , \qquad \tilde{C}(+\infty, t) = 0 \; . \qquad (7)$$

$$\tilde{C}(x,0) = C_m \exp\left[-\frac{(x-R_p)^2}{2\Delta R_p^2} + \frac{v_{BX} x}{2D}\right] \; . \qquad (8)$$

The general solution of the equation (6) for zero boundary conditions (7), i.e. for the case of the interface "the growing layer — epitaxial ambient" completely absorbing impurity atoms, can be obtained by the method of Green's functions

$$\tilde{C}(x,t) = \frac{1}{2\sqrt{\pi D t}} \int_0^{+\infty} \tilde{C}(\xi, 0) G(\xi, x) d\xi \; , \qquad (9)$$

where Green function

$$G(\xi, x) = \exp\left[-\frac{(\xi - x)^2}{4Dt}\right] - \exp\left[-\frac{(\xi + x)^2}{4Dt}\right] \; . \qquad (10)$$

Substituting (8) in (9), we obtain the analytical solution of the equation (6) for the case of ion-implanted impurity redistribution during the epitaxy

$$\tilde{C}(x,t) = \frac{C_m}{2\sqrt{\pi D t}} (I_1 - I_2) \; . \qquad (11)$$

$$I_1 = \int_0^{+\infty} \exp\left[-\frac{(\xi - x)^2}{4Dt} - \frac{(\xi - R_p)^2}{2\Delta R_p^2} + \frac{v_{BX} \xi}{2D}\right] d\xi \qquad (12)$$

$$I_2 = \int_0^{+\infty} \exp\left[-\frac{(\xi + x)^2}{4Dt} - \frac{(\xi - R_p)^2}{2\Delta R_p^2} + \frac{v_{BX} \xi}{2D}\right] d\xi \qquad (13)$$

Now, the problem of calculation of impurity distribution is to integrate the expressions $I_1$ and $I_2$. These integrals can be integrated analytically and are expressed through the exponential functions and error functions erf:

$$I_1 = \frac{1}{u_5} \exp(u_1) \sqrt{\pi} (u_2 + u_3 u_4) \operatorname{erf}\left(\frac{u_4}{2}\right), \qquad (14)$$

$$I_2 = \frac{1}{u_5} \exp(u_6) \sqrt{\pi} \left[1 + \operatorname{erf}(u_5 u_7)\right], \qquad (15)$$

where

$$u_1 = \frac{\Delta R_p^2 v_{BX}(v_{BX} t + 2x)}{4D(\Delta R_p^2 + 2Dt)} \qquad (16)$$

$$- \frac{2D\left[\Delta R_p^2 + x^2 - 2R_p(v_{BX} t + x)\right]}{4D(\Delta R_p^2 + 2Dt)},$$

$$u_2 = 2D R_p t + \Delta R_p^2 (v_{BX} t + x), \qquad (17)$$

$$u_3 = D \Delta R_p^2 t \sqrt{\frac{2}{\Delta R_p^2} + \frac{1}{Dt}}, \qquad (18)$$

$$u_4 = \frac{2DR_p t + \Delta R_p^2 (v_{BX} t + x)}{\Delta R_p \sqrt{Dt(\Delta R_p^2 + 2Dt)}}, \quad (19)$$

$$u_5 = \sqrt{\frac{2}{\Delta R_p^2} + \frac{1}{Dt}}, \quad (20)$$

$$u_6 = \frac{\Delta R_p^2 v_{BX}(v_{BX} t - 2x)}{4D(\Delta R_p^2 + 2Dt)}$$
$$- \frac{2D(R_p^2 - 2R_p v_{BX} t + 2R_p x + x^2)}{4D(\Delta R_p^2 + 2Dt)}, \quad (21)$$

$$u_7 = \frac{2DR_p t + \Delta R_p^2 (v_{BX} t - x)}{2(\Delta R_p^2 + 2Dt)}. \quad (22)$$

**Results of calculations**

The calculations of ion implanted impurity redistribution for two limiting cases, namely, the high and low velocity of epitaxy, are presented on Figure 1 and Figure 2, respectively. It is supposed that the silicon substrate is implanted with boron and then undergoes an epitaxy at the temperature providing a significant impurity diffusion.

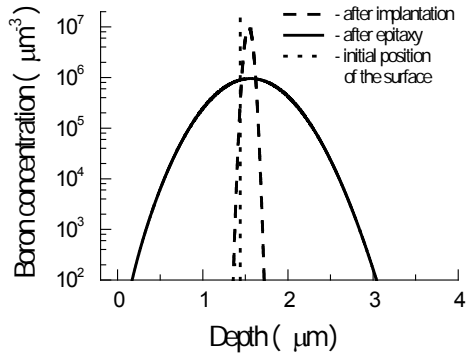

Fig. 1. Calculated profile of ion-implanted boron after epitaxy with high velocity ($V_B$ = -2.4×10$^{-3}$ µm/s)

As can be seen from the Figure 1, at the high velocity of the epitaxy the redistribution of ion-implanted boron is symmetric that is the process of epitaxy has given boron atoms a possibility to diffuse into the left side of semiconductor substrate in the growing epitaxial layer. On the other hand, it is seen from Figure 2 that at low growth velocity the calculated distribution of boron atoms is asymmetrical, as the impurity atoms which have reached the interface of the growing layer due to diffusion process are evaporated.

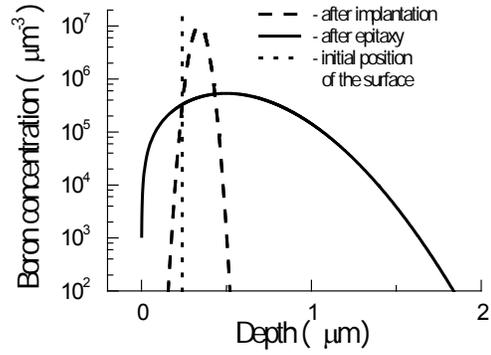

Fig. 2. Calculated profile of ion-implanted boron after epitaxy with low velocity ($V_B$ = -4×10$^{-4}$ µm/s)

The following values of simulation parameters were used:

i) the parameters of initial boron distribution: implantation energy is 30 кэВ; $C_m$ = 1×10$^7$ µm$^{-3}$; $R_p$ = 0.0992 µm; $\Delta R_p$ = 0.0379 µm.

ii) the parameters of diffusion of ion-implanted boron: $D$ = 1×10$^{-4}$ µm$^2$/s; duration of the epitaxy 10 minutes.

**Acknowledgments**
The authors would like to acknowledge I. Berbezier for helpful remarks and advices for future work.

**Conclusion**

The equation of the transient enhanced diffusion of ion-implanted impurity during the epitaxial growing of the layer is presented in the moving coordinate system bound to interface "the growing layer — epitaxial ambient". The equation obtained takes into account the nonuniformity of point defects distribution and drift of mobile species in the field of elastic stresses. The analytical solution of this equation for the case of low fluence ion implantation has been obtained with a view to verify the correctness of the approximate numerical calculations obtained by the codes intended for simulation of impurity diffusion during epitaxy.